\documentclass[conference, 10pt]{IEEEtran}
\IEEEoverridecommandlockouts

\usepackage{color}
\usepackage{times,amsmath,amsfonts,amsthm}
\usepackage{epsfig}
\usepackage{amssymb}
\usepackage{subcaption}
\usepackage{cite}
\usepackage[capitalise]{cleveref}
\usepackage[utf8]{inputenc}
\usepackage{url}

\usepackage{tikz}
\usepackage{pgfplots}
\pgfplotsset{compat=1.17}
\pgfplotstableset{col sep=comma}
\usepackage{moresize}

\input{mysymbol.sty}

\tikzset{every mark/.append style={scale=1.5, solid}, font=\footnotesize}

\pgfplotsset{
    width=1\textwidth,
    legend style={
        font=\ssmall ,  
        inner xsep=1pt,
        inner ysep=1pt,
        nodes={inner sep=1pt}},
    legend cell align=left,
	every axis/.append style={line width=0.5pt},
	every axis plot/.append style={line width=1.25pt},
    every axis y label/.append 
}

\def \bbXo {\bbX_{{\scriptscriptstyle \ccalO}}}

\def \hbCo {\hbC_{{\scriptscriptstyle \ccalO}}}
\def \hbCh {\hbC_{{\scriptscriptstyle \ccalH}}}
\def \hbCoh {\hbC_{{\scriptscriptstyle \ccalO\ccalH}}}
\def \hbCho {\hbC_{{\scriptscriptstyle \ccalH\ccalO}}}

\def \bbSo {\bbS_{{\scriptscriptstyle \ccalO}}}
\def \bbSh {\bbS_{{\scriptscriptstyle \ccalH}}}
\def \bbSoh {\bbS_{{\scriptscriptstyle \ccalO\ccalH}}}
\def \bbSho {\bbS_{{\scriptscriptstyle \ccalH\ccalO}}}
\def \hbSo {\hbS_{{\scriptscriptstyle \ccalO}}}

\title{Joint graph learning from Gaussian observations in the presence of hidden nodes}
\author{Samuel Rey$^{*}$, Madeline Navarro$^{\dag}$, Andrei Buciulea$^{*}$, Santiago Segarra$^{\dag}$, and Antonio G. Marques$^{*}$
	\thanks{Work supported by the Spanish Grants SPGRAPH ( PID2019-105032GB-I00/AEI/10.13039/501100011033), FPU17/04520, EST21/00420, and CAM PEJ-2020-AI/TIC-18964, and the USA NSF award CCF-2008555. $^{*}$Dept. of Signal Theory and Comms., King Juan Carlos University, Madrid, Spain. $^{\dag}$Dept. of ECE, Rice University, Houston, USA. Contact author: antonio.garcia.marques(AT)urjc.es.}}

\begin{document}
\maketitle
\begin{abstract}
Graph learning problems are typically approached by focusing on learning the topology of a single graph when signals from all nodes are available.
However, many contemporary setups involve \emph{multiple} related networks and, moreover, it is often the case that only a subset of nodes is observed while the rest remain \emph{hidden}.
Motivated by this, we propose a \emph{joint graph} learning method that takes into account the presence of \emph{hidden (latent) variables}.
Intuitively, the presence of the hidden nodes renders the inference task ill-posed and challenging to solve, so we overcome this detrimental influence by harnessing the \emph{similarity} of the estimated graphs. 
To that end, we assume that the observed signals are drawn from a Gaussian Markov random field with latent variables and we carefully model the graph similarity among hidden (latent) nodes.
Then, we exploit the structure resulting from the previous considerations to propose a convex optimization problem that solves the joint graph learning task by providing a regularized maximum likelihood estimator.
Finally, we compare the proposed algorithm with different baselines and evaluate its performance over synthetic and real-world graphs.
\end{abstract}
%

\section{Introduction}\label{sec:Introduction}
The rising popularity of graph-based methods to deal with data defined over irregular domains has propelled the development of graph learning algorithms~\cite{mateos2019connecting,sardellitti2019graph,xia2021graph}. 
Indeed, the problem of graph learning, which seeks to learn the topology of a graph from a set of nodal observations, is among the most active research areas of graph signal processing (GSP), a field devoted to the development of tools for processing graph signals~\cite{shuman2013emerging,sandryhaila2013discrete,djuric2018cooperative,rey2022robust}
Fundamental to learning the topology of the graph is assuming that the (statistical) properties of the observed signals depend on the unknown topology, with different assumptions leading to different graph learning methods.
Noteworthy examples of these methods include correlation networks~\cite{kolaczyk2009book}, sparse structural equation models~\cite{cai2013sparse,baingana2014proximal}, graph stationarity models~\cite{segarra2017network,shafipour2020online,navarro2022joint}, smooth (local total variation) models~\cite{kalofolias2016learn,dong2016learning,saboksayr2021accelerated}, and models based on more sophisticated graph-topology priors~\cite{rey2022enhanced,navarro2022jointb}, just to name a few.
Of particular interest for the paper at hand are Gaussian graphical models, a popular and flexible family of graph learning algorithms that builds upon the assumption that the observed
signals are drawn from a Gaussian Markov random field (GMRF)~\cite{lauritzen1996graphical,meinshausen06,friedman2008sparse,ravikumar2011high,egilmez2017graph,kumar2019structured}.

Despite the growing interest aroused by the graph learning problem, the focus is typically placed on learning an \emph{individual} (single) graph under the assumption that observations from \emph{all} nodes are available.
Nonetheless, many contemporary scenarios such as social networks or brain analytics involve multiple related networks, each of them with a subset of available signals.
When several related networks are available, we can harness the graph similarity to boost the performance of the graph learning algorithms~\cite{murase2014multilayer,danaher2014joint,arroyo2021inference,navarro2022joint,rey2022joint}.
Moreover, in many relevant scenarios, we only have access to observations from a subset of nodes, while the rest remain unobserved or hidden.
Intuitively, ignoring the presence of hidden variables will hinder the performance of the graph learning algorithms but, at the same time, accounting for the influence of the hidden nodes renders the graph learning task an ill-posed problem.
This challenging setting has been studied in the context of learning a single graph~\cite{chandrasekaran2012latent,buciulea2019network,chang2019graphical,buciulea2022learning}.
However, modeling the influence of hidden nodes in the context of joint graph learning becomes even more critical since it is unclear how to measure the graph similarity between nodes that remain unobserved.

Motivated by the previous discussion, we propose a convex optimization framework to solve the joint graph learning problem in the presence of hidden variables under the assumption that the observed signals are drawn from a GMRF.
The main motivation is to exploit the known benefits of joint network topology inference methods to improve the performance of the challenging setup resulting from the presence of hidden variables.
Nonetheless, to put forward such an approach we need to carefully model the influence of the hidden nodes in (i) the assumption of GMRF signals that relates the observations and the graph; and (ii) the similarity of the graphs being estimated.
This is achieved by carefully leveraging the structure inherent to the presence of hidden nodes into a regularized maximum likelihood estimator.

Even though the motivation is similar, the resulting algorithm differs substantially from other related works such as~\cite{rey2022joint}.
The main difference lies in the assumption relating the observed signals and the graph topology, which results in modeling the influence of the hidden nodes in a different way, and hence, the structure of the problem is completely different.
Furthermore, the different structure of the problem at hand also requires a new approach for exploiting the graph similarity assumption when hidden nodes are involved.

The remainder of the paper is as follows. \cref{sec:fundamentals} introduces fundamental concepts about GSP and Gaussian graphical models.
\cref{sec:joint_graph_learning} states the joint graph learning problem with hidden nodes and presents the proposed solution.
Finally, \cref{sec:numerical_exp} offers a numerical evaluation of the proposed algorithm and \cref{sec:conclusions} provides some concluding remarks.

\section{Graph learning from Gaussian observations}\label{sec:fundamentals}
\subsection{Fundamentals of GSP}
\vspace{3mm}
\noindent\textbf{Graphs.}
Let $\ccalG = (\ccalV, \ccalE)$ denote an undirected graph described by the set of $N$ nodes $\ccalV$, and the set of links $\ccalE$.
The elements in $\ccalE$ are unordered pairs $(i,j)$ with $i,j \in \ccalV$ and such that $(i,j) \in \ccalE$ implies that nodes $i$ and $j$ are connected.
For any graph $\ccalG$, the adjacency matrix $\bbA \in \reals^{N \times N}$ is a sparse matrix encoding the connectivity of the graph with $A_{ij} = 0$ if and only if $(i,j) \not\in \ccalE$.
Then, the value $A_{ij}$ of the non-zero entries captures the strength of the link between the nodes $i$ and $j$.
  
\vspace{3mm}
\noindent\textbf{Graph signals and graph-shift operator.}
In addition to graphs, we are also concerned with graphs signals, a family of signals defined on the set of nodes $\ccalV$. 
Formally, a graph signal can be denoted as a vector $\bbx = [x_1,...,x_N]^\top \in \reals^N$, where $x_i$ denotes the signal value observed at node $i$.
Because the signal is defined on top of the graph, the key assumption in GSP is that the properties of $\bbx$ are related to the topology of $\ccalG$.
In this sense, if the graph encodes similarity between nodes, then the signal values at two neighboring nodes are expected to be close.
Of particular interest when processing graph signals is the graph-shift operator (GSO), a linear operator applied to graph signals that captures the topology of the graph~\cite{shuman2013emerging}.
The GSO is represented by the matrix $\bbS \in \reals^{N \times N}$ whose entries satisfy that $S_{ij} \neq 0$ only if $i = j$ or $(i,j) \in \ccalE$.
Typical choices for the GSO are the adjacency matrix $\bbA$, the combinatorial graph Laplacian $\bbL := \diag(\bbA\bbone) - \bbA$, or their normalized counterparts~\cite{shuman2013emerging,sandryhaila2013discrete}.

\subsection{Gaussian graphical models}
Arguably, Gaussian graphical models are one of the most commonly used graph learning methods~\cite{lauritzen1996graphical,friedman2008sparse,ravikumar2011high}.
The main assumption adopted by these models is that the observed signals are drawn from a multivariate Gaussian distribution where the zeros in the inverse of the covariance correspond to edges not present in $\ccalE$.
Such a distribution is commonly known as a Gaussian Markov random field (GMRF), and the inverse of the covariance is referred to as the precision matrix.
Upon setting the positive definite precision matrix to be the GSO $\bbS$, the GMRF assumption implies that $\bbx$ is sampled from $\ccalN(\bbzero,\bbS^{-1})$.
In other words, if we denote the covariance of $\bbx$ as $\bbC := \mathbf{E}[\bbx\bbx^\top]$, it follows that the mapping between $\bbC$ and $\bbS$ is given by $\bbC = \bbS^{-1}$.
Then, Gaussian graphical models propose to estimate the graph topology encoded in $\ccalG$ via the regularized maximum likelihood estimator
\begin{alignat}{2}\label{eq:graphical_models}
    \!\!&\!\max_{\bbS} \
    && \;\;\log\det (\bbS) - \tr(\hbC\bbS) - \lambda r(\bbS) \\
    \!\!&\!\mathrm{\;\;s. \;t. } && \;\; \bbS \succeq \bbzero, \nonumber
\end{alignat}
where $\hbC$ denotes the sample covariance matrix, and the term $r(\bbS)$ promotes desired properties on $\bbS$.
Indeed, for the particular case when $r(\bbS)$ is set to $r(\bbS) = \| \bbS \|_1$, the estimator in \eqref{eq:graphical_models} yields the celebrated graphical Lasso algorithm~\cite{friedman2008sparse,danaher2014joint}.

\section{Joint graph learning with hidden nodes}\label{sec:joint_graph_learning}
To formally state the joint graph learning problem in the presence of hidden nodes some definitions are in order.
Let us consider that there are $K$ unknown graphs $\{\ccalG^{(k)}\}_{k=1}^K$, all of them defined over a common set of nodes $\ccalV$ of cardinality $N$.
For the $k$-th graph, we have access to $M_k$ graph signals collected in the matrix $\bbX^{(k)} := [\bbx_1,...,\bbx_{M_k}]^\top \in \reals^{N \times M_k}$.

The presence of hidden nodes implies that only a subset of nodes $\ccalO \subset \ccalV$ of cardinality $O$ is observed.
Meanwhile, the $H = N-O$ remaining nodes collected in the set $\ccalH = \ccalV \setminus \ccalO$ stay unobserved.
Furthermore, consider that the sets $\ccalO$ and $\ccalH$ are the same for the $K$ graphs.
Without loss of generality, let us assume that the nodes in $\ccalO$ correspond to the first nodes in the graph, so the signal values associated with the observed nodes correspond to the first $O$ rows of $\bbX^{(k)}$, which we denote as $\bbXo^{(k)}$. 
Then, let us define the sample covariance matrix of the $k$-th graph as $\hbC^{(k)} := \frac{1}{M_k}\bbX^{(k)}(\bbX^{(k)})^\top$. 
The distinction between observed and hidden nodes endows the GSO and the sampled covariance matrix with the following block structure
\begin{equation}\label{eq:block_structure}
    \bbS =
    \begin{bmatrix}
    \bbSo & \bbSoh  \\
    \bbSho & \bbSh
    \end{bmatrix}\!, \;
    \hbC =
    \begin{bmatrix}
    \hbCo & \hbCoh  \\
    \hbCho & \hbCh
    \end{bmatrix}. \;
\end{equation}
The $O \times O$ submatrices $\bbSo^{(k)}$ and $\hbCo^{(k)}$ respectively denote the block of the GSO capturing the connection between observed nodes and the block of the sample covariance of the signal values at the observed nodes (i.e., the sample covariance of $\bbXo$).
On the other hand, the remaining blocks of $\bbS$
and $\hbC$ involve edges and covariances between hidden nodes.

Based on the previous definitions, the problem of joint graph learning in the presence of hidden nodes is formally stated next.
\begin{problem}\label{p:joint_gl_hidden}
    Find the matrices $\{\bbSo^{(k)}\}_{k=1}^K$ encoding the connectivity between the observed nodes $\ccalO$ for all of the $K$ graphs, given the $O \times M_k$ matrices $\{\bbXo^{(k)}\}_{k=1}^K$ collecting the signal values at the observed nodes under the assumptions that: \\
    \textbf{(AS1)} The number of observed nodes is considerably larger than the number of hidden nodes, i.e., $O \gg H$.  \\
    \textbf{(AS2)} The columns of $\bbX^{(k)}$ are independent realizations of a zero mean multivariate Gaussian distribution $\ccalN(\bbzero, (\bbS^{(k)})^{-1})$. \\
    \textbf{(AS3)} The distance between the $K$ graphs is small according to  some metric $d(\bbS^{(k)},\bbS^{(k')})$.
\end{problem}

Learning the topology encoded in the GSOs $\{\bbS^{(k)}_{\ccalO}\}_{k=1}^K$ while accounting for the influence of hidden nodes is an ill-conditioned problem since there are no observations from the nodes in $\ccalH$.
Therefore, \textbf{(AS1)} ensures the tractability of the problem by assuming that most of the nodes are observed.
Second, \textbf{(AS2)} establishes a connection between the unknown graph $\ccalG^{(k)}$ and the signals collected in $\bbX^{(k)}$ via a GMRF model.
However, although this assumption involves the whole matrices $\bbX^{(k)}$, only the submatrices $\bbXo^{(k)}$ are observed.
A similar problem arises with \textbf{(AS3)}.
The last assumption guarantees that the $K$ graphs are closely related so we can harness this relation in a joint graph learning algorithm.
Nonetheless, how to exploit the similarity between subgraphs whose nodes are not being observed is not a trivial question.

To address these issues, in the following, we leverage the block structure introduced in \eqref{eq:block_structure} to model how the presence of the hidden nodes carries over to the assumptions \textbf{(AS1)} and \textbf{(AS2)}, and then we leverage the resulting relations to solve the joint graph learning problem with hidden variables via an optimization problem.

\subsection{Joint graph learning as an optimization problem}
Let us start by modeling how the presence of hidden variables impacts the GMRF assumption \textbf{(AS2)}, which is fundamental for the graph learning problem at hand.
Recall that the GMRF assumption implies that the mapping between the covariance matrix and the GSO is given by $\bbC^{(k)}=(\bbS^{(k)})^{-1}$.
Then, due to the presence of hidden nodes, we only have access to the observed sampled covariance matrix $\hbCo^{(k)}$, and hence, we are interested in establishing a relation between the matrices $\bbSo^{(k)}$ and $\hbCo^{(k)}$.
To that end, similar to \cite{chandrasekaran2012latent,buciulea2022learning}, we leverage the block structure of $\hbCo^{(k)}$ and $\bbSo^{(k)}$, and employ the Schur complement to obtain the following expression
\begin{equation}
    (\hbCo^{(k)})^{-1} = \bbSo^{(k)} - \bbSoh^{(k)}(\bbSh^{(k)})^{-1}\bbSho^{(k)} = \bbSo^{(k)} - \bbP^{(k)}.
\end{equation}
Since all the blocks of the GSO associated with hidden nodes are unknown, we \emph{lift} the problem by defining the matrices $\bbP^{(k)} :=  \bbSoh^{(k)}(\bbSh^{(k)})^{-1}\bbSho^{(k)} \in \reals^{O \times O}$.
These matrices capture the influence of the hidden nodes and, since they involve the product of the $H \times H$ matrices $\bbSh^{(k)}$, from \textbf{(AS1)} it follows that the matrices $\bbP^{(k)}$ are low-rank matrices such that $\rank(\bbP^{(k)}) \leq H$.
Furthermore, we can leverage these matrices to harness the graph similarity \textbf{(AS3)} between hidden nodes, as we will explain in more detail later.

Then, a reasonable approach to tackle \cref{p:joint_gl_hidden} is to modify the regularized maximum likelihood estimator from \eqref{eq:graphical_models} based on the aforementioned considerations.
This result in the non-convex optimization problem
\begin{alignat}{2}\label{eq:joint_hidden_gl_noncvx}
    \!\!&\!
    \min_{\{\bbSo^{(k)},\bbP^{(k)}\}_{k=1}^K} \
    && 
    \!\!\sum_{k=1}^K \tr\left((\bbSo^{(k)}\!-\!\bbP^{(k)})\hbCo^{(k)}\right) \!-\! \log\det(\bbSo^{(k)}\!-\!\bbP^{(k)}) \nonumber \\
     \!\!&\!  &&  
    \!\! + \rho_k\|\bbSo^{(k)}\|_0 \nonumber \\
    \!\!&\!  &&
    \!\! + \sum_{k < k'}\rho_{kk'} d_S(\bbSo^{(k)}, \bbSo^{(k')}) + \beta_{kk'} d_P(\bbP^{(k)}, \bbP^{(k')}) \nonumber \\
    \!\!&\!
    \mathrm{\;\;s. \;t. } 
    &&
    \bbP^{(k)} \succcurlyeq 0;\;\; \bbS^{(k)}-\bbP^{(k)} \succ 0;\;\; \bbSo^{(k)} \in \ccalS, \;\; \nonumber \\
    \!\!&\!  && \rank(\bbP^{(k)}) \leq H.
\end{alignat}
First, note that the maximum likelihood terms involve the expression $\bbSo^{(k)}-\bbP^{(k)}$ to account for the influence of hidden nodes in assumption \textbf{(AS2)}.
Second, the rank constraint $\rank(\bbP^{(k)}) \leq H$ captures the fact that the matrices $\bbP^{(k)}$ involve the multiplication by $H \times H$ matrices.
Then, the $\ell_0$ promotes sparsity on the estimated GSOs and the set $\ccalS$ represents the set of admissible GSOs; see~\cite{segarra2017network}.

Regarding \textbf{(AS3)}, to harness the similarity of the $K$ graphs on the whole GSOs (instead of only on the observed blocks), the objective function in \eqref{eq:joint_hidden_gl_noncvx} includes two distances.
Inspired by other standard joint inference approaches~\cite{danaher2014joint}, we capture the similarity between the observed GSOs via the function $d_S(\cdot,\cdot)$.
Then, the function $d_P(\cdot,\cdot)$ captures the graph similarity between the unobserved blocks of the GSOs by minimizing the distance between the matrices $\bbP^{(k)}$.
This important distinction allows us to incorporate additional structure reducing the degrees of freedom and rendering the problem more manageable.

While there are several ways of measuring graph similarity, in this work we focus on the commonly used assumption that the $K$ graphs present a similar support~\cite{danaher2014joint,navarro2022joint}.
The influence on the observed nodes is straightforward  since such an assumption implies that the sparsity pattern of the matrices $\bbSo^{(k)}$ is similar, so we can set
\begin{equation}
    d_S(\bbSo^{(k)},\bbSo^{(k')}) = \| \bbSo^{(k)} - \bbSo^{(k')} \|_0.
\end{equation}
On the other hand, to select an appropriate function $d_P(\cdot,\cdot)$ we recall that $\bbP^{(k)} = \bbSoh^{(k)}(\bbSh^{(k)})^{-1}\bbSho^{(k)}$.
From this expression, it follows that the submatrices $\bbSoh^{(k)}$ and $\bbSho^{(k)}$ determine the sparsity pattern of the matrices $\bbP^{(k)}$, and hence, the graph similarity assumption also implies that the sparsity pattern of the matrices $\bbP^{(k)}$ has to be similar across the different graphs.
Therefore, we select the function
\begin{equation}
    d_P(\bbP^{(k)},\bbP^{(k')}) = \| \bbP^{(k)} - \bbP^{(k')} \|_0.
\end{equation}
Although the resulting distance is simple, it builds upon the known structure of the matrices $\{\bbP^{(k)}\}_{k=1}^K$ to harness the similarity assumption over hidden nodes.

Once the graph similarity distance is set, the final ingredient to approach the joint graph learning problem in the presence of hidden nodes is to deal with the non-convexity of \eqref{eq:joint_hidden_gl_noncvx}.
The non-convexity of the optimization problem arises from the presence of the $\ell_0$ norm and the rank constraint.
Following the classical approach in the literature, we replace the $\ell_0$ norm by its convex surrogate the $\ell_1$ norm, and instead of setting a maximum value for the rank of each $\bbP^{(k)}$, we minimize its nuclear norm.
This result in the following convex optimization problem
\begin{alignat}{2}\label{eq:joint_hidden_gl}
    \!\!&\!
    \min_{\{\bbSo^{(k)},\bbP^{(k)}\}_{k=1}^K} \
    && 
    \!\!\sum_{k=1}^K \tr\left((\bbSo^{(k)}\!-\!\bbP^{(k)})\hbCo^{(k)}\right) \!-\! \log\det(\bbSo^{(k)}\!-\!\bbP^{(k)}) \nonumber \\
     \!\!&\!  &&  
    \!\! + \rho_k\|\bbSo^{(k)}\|_{1} + \beta_k\|\bbP^{(k)}\|_* \nonumber \\
    \!\!&\!  &&
    \!\! + \sum_{k < k'}\rho_{kk'}\|\bbSo^{(k)}\!-\!\bbSo^{(k')}\|_1 \!+\! \beta_{kk'} \|\bbP^{(k)}\!-\!\bbP^{(k')}\|_1 \nonumber \\
    \!\!&\!
    \mathrm{\;\;s. \;t. } 
    &&
    \bbP^{(k)} \succcurlyeq 0;\;\; \bbS^{(k)}-\bbP^{(k)} \succ 0;\;\; \bbSo^{(k)} \in \ccalS, \;\;
\end{alignat}
where $\| \bbSo^{(k)} \|_1$ denotes the $\ell_1$ norm of the vectorized form of $\bbSo^{(k)}$, and $\| \bbP^{(k)} \|_*$ denotes the nuclear norm of $\bbP^{(k)}$. 


\section{Numerical results}\label{sec:numerical_exp}
In this section, we evaluate the performance of the proposed algorithms over both synthetic and real-world graphs.
The code related to the implementation of our graph learning algorithm and the code for all the simulations presented in this paper is available in GitHub\footnote{\url{https://github.com/reysam93/hidden_joint_gaussian_inf}}.

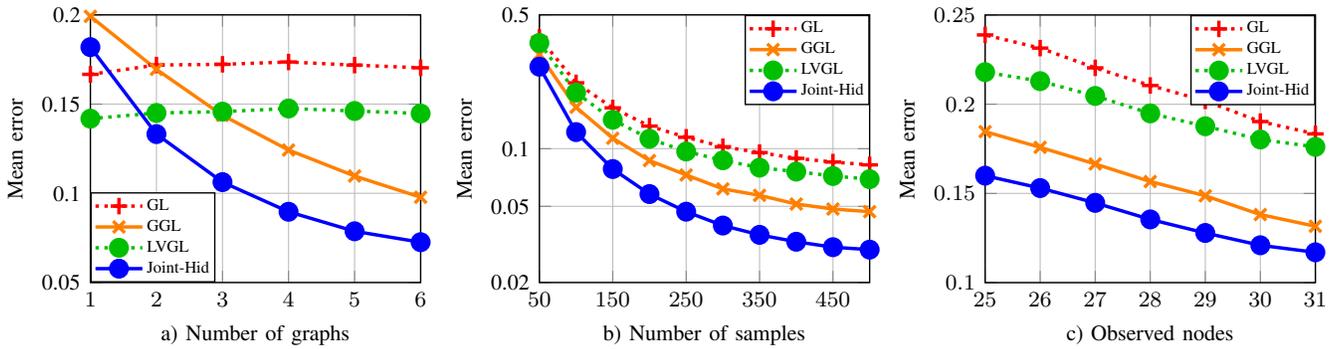
\begin{figure*}[!t]
	\centering
	\begin{subfigure}{0.32\textwidth}
		\centering
		 \begin{tikzpicture}[baseline,scale=1.04]
\begin{axis}[
    xlabel={a) Number of graphs},
    xmin={1},
    xmax={6},
    xtick={1,2,...,6},
    ylabel={Mean error},
    ymin={0.05},
    ymax={0.2},
    ytick={0.05, 0.1, 0.15, 0.2},
    yticklabels={0.05, 0.1, 0.15, 0.2},
    grid=major,
    legend style={
        at={(0,0)},
        anchor=south west}]
    
    \pgfplotstableread{data/joint_inf_v2.csv}\errtable
    
    \addplot[red, dotted, mark=+] table [x=xaxis, y=GL] {\errtable};
    \addplot[orange, mark=x] table [x=xaxis, y=GGL] {\errtable};
    \addplot[green!75!black, dotted, mark=*] table [x=xaxis, y=LVGL] {\errtable};
    \addplot[blue, mark=*] table [x=xaxis, y=Joint] {\errtable}; 
    
    \legend{GL, GGL, LVGL, Joint-Hid}

\end{axis}
\end{tikzpicture}
	\end{subfigure}
	\begin{subfigure}{0.32\textwidth}
		\centering
        \begin{tikzpicture}[baseline,scale=1.04]
\begin{semilogyaxis}[
    xlabel={b) Number of samples},
    xmin={50},
    xmax={500},
    xtick={50,150,...,450},
    ylabel={Mean error},
    ymin={0.02},
    ymax={0.5},
    ytick={0.02, 0.05, 0.1, 0.5},
    yticklabels={0.02, 0.05, 0.1, 0.5},
    grid=major,
    legend style={
        at={(1,1)},
        anchor=north east}]
    
    \pgfplotstableread{data/samples_inf_v4.csv}\errtable
    
    \addplot[red, dotted, mark=+] table [x=xaxis, y=GL] {\errtable};
    \addplot[orange, mark=x] table [x=xaxis, y=GGL] {\errtable};
    \addplot[green!75!black, dotted, mark=*] table [x=xaxis, y=LVGL] {\errtable};
    \addplot[blue, mark=*] table [x=xaxis, y=Joint] {\errtable}; 
    
    \legend{GL, GGL, LVGL, Joint-Hid}

\end{semilogyaxis}
\end{tikzpicture}
	\end{subfigure}
	\begin{subfigure}{0.32\textwidth}
		\centering
		\begin{tikzpicture}[baseline,scale=1.04]
\begin{axis}[
    xlabel={c) Observed nodes},
    xmin={25},
    xmax={31},
    xtick={25,26,...,31},
    ylabel={Mean error},
    ymin={0.1},
    ymax={0.25},
    grid=major,
    legend style={
        at={(1,1)},
        anchor=north east}]
    
    \pgfplotstableread{data/students_hidden.csv}\errtable
    
    \addplot[red, dotted, mark=+] table [x=xaxis, y=GL] {\errtable};
    \addplot[orange, mark=x] table [x=xaxis, y=GGL] {\errtable};
    \addplot[green!75!black, dotted, mark=*] table [x=xaxis, y=LVGL] {\errtable};
    \addplot[blue, mark=*] table [x=xaxis, y=Joint] {\errtable}; 
    
    \legend{GL, GGL, LVGL, Joint-Hid}

\end{axis}
\end{tikzpicture}
	\end{subfigure}
	\caption{Mean error of the estimated matrices $\hbS^{(k)}$ through several algorithms and in different types of graphs. The error is computed according to \eqref{eq:error_metric} and we report the mean of 100 independent realizations. a)~Graphs being estimated are ER graphs with $N=20$ nodes and $p=0.15$. b)~Graphs are drawn from an SW random graph model with $N=20$ nodes, 3 neighbors per node, and a rewiring probability of $p=0.15$. c)~Graphs are obtained from the students of Ljubljana dataset.} \label{fig:experiments}
\end{figure*}

The results of the experiments conducted are shown in \cref{fig:experiments}, where we compare the estimated graphs obtained with our algorithm in \eqref{eq:joint_hidden_gl}, denoted as ``Joint-Hid'' in the legend, with other popular graphical models.
The algorithms considered as baselines are: (i)~the graphical Lasso algorithm, denoted as ``GL''~\cite{friedman2008sparse}; (ii)~group graphical Lasso, a variant of the graphical Lasso algorithm modified to jointly estimate several related graphs that is denoted as ``GGL''~\cite{danaher2014joint}; and (iii) latent-variable graphical Lasso, a variant of graphical Lasso that accounts for the presence of hidden nodes that is denoted as ``LVGL''~\cite{chandrasekaran2012latent}.   
Then, to prevent additional errors associated with the scale ambiguity that often appears in graph learning algorithms, we consider the following mean error metric
\begin{equation}\label{eq:error_metric}
    \frac{1}{K}\sum_{k=1}^K \left\| \frac{\hbSo^{(k)}}{\|\hbSo^{(k)}\|_F} - \frac{\bbSo^{(k)}}{\|\bbSo^{(k)}\|_F} \right\|_F^2,
\end{equation}
where $\hbSo^{(k)}$ and $\bbSo^{(k)}$ respectively denote the estimated and the true observed GSO.
The error reported in the experiments is the average of 100 independent realizations.
The different experiments and the obtained results are detailed next.

\vspace{3mm}
\noindent\textbf{Test case 1.} 
We start by assessing the benefits of implementing a joint graph learning approach.
The results are depicted in \cref{fig:experiments}a, where the x-axis shows the number of related graphs that need to be estimated and the y-axis the error computed as in \eqref{eq:error_metric}.
The first graph $\ccalG^{(1)}$ is generated as an Erd\H{o}s-Rényi (ER) with $N=20$ nodes and a link probability of $p=0.2$ and the remaining $K-1$ graphs are generated by randomly rewiring a fix number of edges. 
Among the 20 nodes, $H=2$ nodes are selected as hidden nodes according to a uniform distribution so only $O=18$ nodes are observed.
Then, for each graph we create $M_k = 200$ graph signals drawn from a multivariate Gaussian distribution $\ccalN(\bbzero, (\bbS^{(k)})^{-1})$.
From the results in \cref{fig:experiments}a, we observe that the error of ``Joint-Hid'' and ``GGL'' drops as the number of related graphs increases, showcasing the benefit of the joint graph learning approach.
In addition, we also observe the importance of accounting for the presence of hidden variables since ``LVGL'' outperforms both ``GL'' and ``GGL'' when the number of graphs being estimated is small.
Finally, we remark that the ``Joint-Hid'' algorithm outperforms the other alternatives, which was expected since it considers both the influence of hidden nodes and the joint estimation of the graphs.

\vspace{3mm}
\noindent\textbf{Test case 2.} 
The second experiment compares the performance of different algorithms as the number of observed signals increases.
The results are depicted in \cref{fig:experiments}b where the x-axis shows the number of samples $M_k$, which is the same for the $K=4$ graphs.
In this case, the graphs considered are small world (SW) graphs~\cite{watts1998collective} with $N=20$ nodes, each node is connected to 3 neighbors, and the rewiring probability is $p=0.15$.
From the 20 nodes, $O=18$ are observed nodes while the remaining $H=2$ stay hidden.
The results in \cref{fig:experiments}b show that the performance of all the algorithms is similar when the number of signals is small but, as more signals are available, the proposed ``Joint-Hid'' method swiftly outperforms the alternatives.
On the opposite side, the ``GL'' method requires a larger amount of signals to obtain an error below $0.1$.
As a result, this experiment highlights that following a joint approach that takes into account the hidden variables allows more efficient usage of the available signals, rendering a smaller error with fewer observed signals.

\vspace{3mm}
\noindent\textbf{Test case 3.}
Finally, we close the numerical evaluation by studying the influence of the number of hidden nodes.
In this case, we employ $K=4$ real-world graphs extracted from the network of students of the University of Ljubljana dataset\footnote{The original data can be found at {\scriptsize \url{http://vladowiki.fmf.uni-lj.si/doku.php?id=pajek:data:pajek:students}}}.
All the graphs are composed of the same set of 32 nodes, which represent students, and the different networks capture different types of interactions among the students.
The graph signals have been generated according to a GMRF as in the previous experiments.
The error as the number of observed nodes increases (and hence, the number of hidden nodes becomes smaller), is illustrated in \cref{fig:experiments}c.
Looking at the figure, it is clear that the proposed ``Joint-Hid'' algorithm outperforms the other alternatives.
Moreover, we note that, as the number of hidden nodes gets closer to 0, the performances of ``GL'' and ``LVGL'' seems to converge.
The same happens with ``GGL'' and ``Joint-Hid''.
Interestingly, this behavior suggests that no evident prejudice follows from accounting for hidden nodes when all the nodes are being observed.

\section{Conclusions}\label{sec:conclusions}
In this paper, we approached the challenging task of joint graph learning of several closely related graphs in the presence of hidden nodes.
To establish a relation between the observed signals and the unknown graphs, we assumed that the signals were drawn from a GMRF model, i.e., $\bbX^{(k)} \sim \ccalN(\bbzero, (\bbS^{(k)})^{-1})$.
Because the presence of hidden nodes renders the problem ill-conditioned, we further assumed that the number of observed nodes is much larger than the number of hidden nodes ($O \gg H$) to ensure the tractability of the problem.
Then, we modified the GMRF assumption to account for the hidden nodes by exploiting the block structure of $\bbS^{(k)}$ and $\hbC^{(k)}$, resulting in the definition of the low-rank matrices $\bbP^{(k)}$.
Moreover, after studying the definition of these matrices, we arrived at the conclusion that we could harness the graph similarity between blocks of $\bbS^{(k)}$ associated with hidden nodes by estimating matrices $\bbP^{(k)}$ with a similar sparsity pattern.
This observation allowed us to leverage the graph similarity on the \emph{whole graphs}.
Finally, we presented a convex algorithm employing the $\ell_1$ norm and the nuclear norm as convex surrogates for the $\ell_0$ norm and the matrix rank.
The performance of the proposed algorithm was compared with other popular Gaussian graphical models over synthetic and real-world graphs.

\bibliographystyle{IEEEbib}
\bibliography{myIEEEabrv,biblio}

\begin{thebibliography}{10}

\bibitem{mateos2019connecting}
G.~Mateos, S.~Segarra, A.~G. Marques, and A.~Ribeiro,
\newblock ``Connecting the dots: Identifying network structure via graph signal
  processing,''
\newblock {\em IEEE Signal Process. Mag.}, vol. 36, no. 3, pp. 16--43, 2019.

\bibitem{sardellitti2019graph}
S.~Sardellitti, S.~Barbarossa, and P.~Di~Lorenzo,
\newblock ``Graph topology inference based on sparsifying transform learning,''
\newblock {\em IEEE Trans. Signal Process.}, vol. 67, no. 7, pp. 1712--1727,
  2019.

\bibitem{xia2021graph}
F.~Xia, K.~Sun, S.~Yu, A.~Aziz, L.~Wan, S.~Pan, and H.~Liu,
\newblock ``Graph learning: A survey,''
\newblock {\em IEEE Trans. Artif. Intell.}, vol. 2, no. 2, pp. 109--127, 2021.

\bibitem{shuman2013emerging}
D.I. Shuman, S.K. Narang, P.~Frossard, A.~Ortega, and P.~Vandergheynst,
\newblock ``The emerging field of signal processing on graphs: Extending
  high-dimensional data analysis to networks and other irregular domains,''
\newblock {\em IEEE Signal Process. Mag.}, vol. 30, no. 3, pp. 83--98, 2013.

\bibitem{sandryhaila2013discrete}
A.~Sandryhaila and J.~M.~F. Moura,
\newblock ``Discrete signal processing on graphs,''
\newblock {\em IEEE Trans. Signal Process.}, vol. 61, no. 7, pp. 1644--1656,
  2013.

\bibitem{djuric2018cooperative}
P.~Djuric and C.~Richard,
\newblock {\em Cooperative and Graph Signal Processing: Principles and
  Applications},
\newblock Academic Press, 2018.

\bibitem{rey2022robust}
S.~Rey, V.~M. Tenorio, and A.~G. Marques,
\newblock ``Robust graph filter identification and graph denoising from signal
  observations,''
\newblock {\em arXiv preprint arXiv:2210.08488}, 2022.

\bibitem{kolaczyk2009book}
E.~D. Kolaczyk,
\newblock {\em Statistical Analysis of Network Data: Methods and Models},
\newblock Springer, New York, NY, 2009.

\bibitem{cai2013sparse}
X.~Cai, J.~A. Bazerque, and G.~B. Giannakis,
\newblock ``Sparse structural equation modeling for inference of gene
  regulatory networks exploiting genetic perturbations,''
\newblock {\em PLoS, Comput. Biology}, June 2013.

\bibitem{baingana2014proximal}
B.~Baingana, G.~Mateos, and G.~B. Giannakis,
\newblock ``Proximal-gradient algorithms for tracking cascades over social
  networks,''
\newblock {\em IEEE J. Sel. Topics Signal Process.}, vol. 8, pp. 563--575, Aug.
  2014.

\bibitem{segarra2017network}
S.~Segarra, A.~G. Marques, G.~Mateos, and A.~Ribeiro,
\newblock ``Network topology inference from spectral templates,''
\newblock {\em IEEE Trans. Signal Inf. Process. Netw.}, vol. 3, no. 3, pp.
  467--483, 2017.

\bibitem{shafipour2020online}
R.~Shafipour and G.~Mateos,
\newblock ``Online topology inference from streaming stationary graph signals
  with partial connectivity information,''
\newblock {\em Algorithms}, vol. 13, no. 9, pp. 228, 2020.

\bibitem{navarro2022joint}
M.~Navarro, Y.~Wang, A.~G. Marques, C.~Uhler, and S.~Segarra,
\newblock ``Joint inference of multiple graphs from matrix polynomials,''
\newblock {\em J. Mach. Learn. Res.}, vol. 23, no. 76, pp. 1--35, 2022.

\bibitem{kalofolias2016learn}
V.~Kalofolias,
\newblock ``How to learn a graph from smooth signals,''
\newblock in {\em Intl. Conf. Artif. Intel. Statist. (AISTATS)}. J. Mach.
  Learn. Res., 2016, pp. 920--929.

\bibitem{dong2016learning}
X.~Dong, D.~Thanou, P.~Frossard, and P.~Vandergheynst,
\newblock ``Learning {L}aplacian matrix in smooth graph signal
  representations,''
\newblock {\em IEEE Trans. Signal Process.}, vol. 64, no. 23, pp. 6160--6173,
  2016.

\bibitem{saboksayr2021accelerated}
S.~S. Saboksayr and G.~Mateos,
\newblock ``Accelerated graph learning from smooth signals,''
\newblock {\em IEEE Signal Process. Lett.}, vol. 28, pp. 2192--2196, 2021.

\bibitem{rey2022enhanced}
S.~Rey, T.~M. Roddenberry, S.~Segarra, and A.~G. Marques,
\newblock ``Enhanced graph-learning schemes driven by similar distributions of
  motifs,''
\newblock {\em arXiv preprint arXiv:2207.04747}, 2022.

\bibitem{navarro2022jointb}
M.~Navarro and S.~Segarra,
\newblock ``Joint network topology inference via a shared graphon model,''
\newblock {\em IEEE Trans. Signal Process.}, 2022.

\bibitem{lauritzen1996graphical}
S.~L. Lauritzen,
\newblock {\em Graphical Models}, vol.~17,
\newblock Clarendon Press, 1996.

\bibitem{meinshausen06}
N.~Meinshausen and P.~Buhlmann,
\newblock ``High-dimensional graphs and variable selection with the lasso,''
\newblock {\em Ann. Statist.}, vol. 34, pp. 1436--1462, 2006.

\bibitem{friedman2008sparse}
J.~Friedman, T.~Hastie, and R.~Tibshirani,
\newblock ``Sparse inverse covariance estimation with the graphical lasso,''
\newblock {\em Biostatistics}, vol. 9, no. 3, pp. 432--441, 2008.

\bibitem{ravikumar2011high}
P.~Ravikumar, M.~J. Wainwright, G.~Raskutti, and B.~Yu,
\newblock ``High-dimensional covariance estimation by minimizing
  $\ell$1-penalized log-determinant divergence,''
\newblock {\em Electron. J. Statistics}, vol. 5, pp. 935--980, 2011.

\bibitem{egilmez2017graph}
H.~E. Egilmez, E.~Pavez, and A.~Ortega,
\newblock ``Graph learning from data under {L}aplacian and structural
  constraints,''
\newblock {\em IEEE J. Sel. Topics Signal Process.}, vol. 11, no. 6, pp.
  825--841, 2017.

\bibitem{kumar2019structured}
S.~Kumar, J.~Ying, J.~Cardoso, and D.~Palomar,
\newblock ``Structured graph learning via {L}aplacian spectral constraints,''
\newblock {\em Advances Neural Inf. Process. Syst.}, vol. 32, 2019.

\bibitem{murase2014multilayer}
Y.~Murase, J.~T{\"o}r{\"o}k, H.~H. Jo, K.~Kaski, and J.~Kert{\'e}sz,
\newblock ``Multilayer weighted social network model,''
\newblock {\em Physical Review E}, vol. 90, no. 5, pp. 052810, 2014.

\bibitem{danaher2014joint}
P.~Danaher, P.~Wang, and D.~M. Witten,
\newblock ``The joint graphical lasso for inverse covariance estimation across
  multiple classes,''
\newblock {\em J. Roy. Statistical Soc.: Ser. B (Statistical Methodology)},
  vol. 76, no. 2, pp. 373--397, 2014.

\bibitem{arroyo2021inference}
J.~Arroyo, A.~Athreya, J.~Cape, G.~Chen, C.~E. Priebe, and J.~T. Vogelstein,
\newblock ``Inference for multiple heterogeneous networks with a common
  invariant subspace,''
\newblock {\em J. Mach. Learn. Res.}, vol. 22, no. 142, pp. 1--49, 2021.

\bibitem{rey2022joint}
S.~Rey, A.~Buciulea, M.~Navarro, S.~Segarra, and A.~G. Marques,
\newblock ``Joint inference of multiple graphs with hidden variables from
  stationary graph signals,''
\newblock in {\em IEEE Int. Conf. Acoustics, Speech Signal Process. (ICASSP)}.
  IEEE, 2022, pp. 5817--5821.

\bibitem{chandrasekaran2012latent}
V.~Chandrasekaran, P.~A. Parrilo, and A.~S. Willsky,
\newblock ``Latent variable graphical model selection via convex
  optimization,''
\newblock {\em Ann. Statist.}, vol. 40, no. 4, pp. 1935--1967, 2012.

\bibitem{buciulea2019network}
A.~Buciulea, S.~Rey, C.~Cabrera, and A.~G Marques,
\newblock ``Network reconstruction from graph-stationary signals with hidden
  variables,''
\newblock in {\em Conf. Signals, Syst., Computers (Asilomar)}. IEEE, 2019, pp.
  56--60.

\bibitem{chang2019graphical}
A.~Chang, T.~Yao, and G.~I. Allen,
\newblock ``Graphical models and dynamic latent factors for modeling functional
  brain connectivity,''
\newblock {\em 2019 IEEE Data Science Wrksp. (DSW)}, pp. 57--63, 2019.

\bibitem{buciulea2022learning}
A.~Buciulea, S.~Rey, and A.~G. Marques,
\newblock ``Learning graphs from smooth and graph-stationary signals with
  hidden variables,''
\newblock {\em IEEE Trans. Signal Inf. Process. Netw.}, 2022.

\bibitem{watts1998collective}
D.~J. Watts and S.~H. Strogatz,
\newblock ``Collective dynamics of ‘small-world’ networks,''
\newblock {\em Nature}, vol. 393, no. 6684, pp. 440--442, 1998.

\end{thebibliography}

\end{document}